\documentclass{ws-mpla}
\usepackage[super]{cite}
\usepackage{graphicx}
\begin{document}

\markboth{M.Blasone \& L.Smaldone}
{A note on oscillating neutrino states in quantum field theory}
\newcommand{\be}{\begin{equation}}\newcommand{\ee}{\end{equation}}
\newcommand{\bea}{\begin{eqnarray}}\newcommand{\eea}{\end{eqnarray}}
\newcommand{\brr}{\begin{array}}\newcommand{\err}{\end{array}}
\newcommand{\bit}{\begin{itemize}}\newcommand{\eit}{\end{itemize}}
\newcommand{\ben}{\begin{enumerate}}\newcommand{\een}{\end{enumerate}}
\newcommand{\bib}{\bibitem}
\newcommand{\bbm}{\begin{bmatrix}}\newcommand{\ebm}{\end{bmatrix}}
\newcommand{\ba}{\begin{array}}
\newcommand{\ea}{\end{array}}
\newcommand{\G}{\textbf}
\newcommand{\C}{\textit}
\newtheorem{mydef}{Definition}
\newtheorem{Lemma}{Lemma}
\newcommand{\bd}{\begin{mydef}} \newcommand{\ed}{\end{mydef}}
\newcommand{\bthe}{\begin{theorem}} \newcommand{\ethe}{\end{theorem}}
\newcommand{\ble}{\begin{Lemma}} \newcommand{\ele}{\end{Lemma}}
\newcommand{\ul}{\underline}
\newcommand{\mc}{\mathcal}
\newcommand{\mr}{\mathrm}
\newcommand{\dr}{\mathrm{d}}
\def\ha{\frac{1}{2}}
\def\lag{\mathcal{L}}
\def\intx{\int \!\!\mathrm{d}^3 {\G x}}
\def\intk{\int \!\!\mathrm{d}^3 {\G k}}
\def\tr{\mathrm{tr}}
\def\Tr{\mathrm{Tr}}
\def\ph{\varphi}
\def\Dia{\Diamond}\def\lab{\label}\def\lan{\langle}\def\lar{\leftarrow}
\def\lf{\left}\def\lrar{\leftrightarrow}
\def\Lrar{\Leftrightarrow}\def\noi{\noindent}
\def\non{\nonumber}\def\ot{\times}\def\pa{\partial}\def\ran{\rangle}
\def\rar{\rightarrow}\def\Rar{\Rightarrow}
\def\ri{\right}\def\ti{\tilde}\def\we{\wedge}\def\wti{\widetilde}
\def\al{\alpha}\def\bt{\beta}\def\ga{\gamma}\def\Ga{\Gamma}
\def\de{\delta}\def\De{\Delta}\def\ep{\epsilon}
\def\ze{\zeta}\def\te{\theta}\def\ka{\kappa}
\def\la{\lambda}\def\La{\Lambda}\def\si{\sigma}\def\Si{\Sigma}
\def\om{\omega}\def\Om{\Omega}
\def\AB{{_{A,B}}}\newcommand{\mlab}[1]{\label{#1}}
\def\CP{{_{C\!P}}}\def\T{{_{T}}}
\def\AB{{_{A,B}}}\def\mass{{_{1,2}}}
\def\flav{{e,\mu}}\def\1{{_{1}}}\def\2{{_{2}}}
\def\bp{{\bf {p}}}\def\bk{{\bf {k}}}\def\br{{\bf {r}}}\def\bx{{\bf {x}}}
\def\by{{\bf {y}}}\def\bl{{\bf {l}}}\def\bq{{\bf {q}}}\def\bj{{\bf {j}}}
\def \ak{\alpha^r_{{\bf k},e}(0)}\def \akd{\alpha^{r\dag}_{{\bf k},e}(0)}
\def\ap{\alpha^s_{{\bf p},e}(0)}\def\apd{\alpha^{s\dag}_{{\bf p},e}(0)}
\def\br{{\bf{r}}}\def\bI{{\bf{I}}}
\def\fourint{\int\!\!\!\int\!\!\!\int\!\!\!\int}\def\threeint{\int\!\!\!\int\!\!\!\int}
\def\twoint{\int\!\!\!\int}
\def\CKM{\tiny CKM}
\def\Qo{{_{Q\!,1}}}\def\Qt{{_{Q\!,2}}}\def\ko{{_{k\!,1}}}\def\kt{{_{k\!,2}}}
\newcommand{\ide}{1\hspace{-1mm}{\rm I}}
\newcommand{\noH}{:\;\!\!\;\!\!:H:\;\!\!\;\!\!:}
\def\noHe0{:\;\!\!\;\!\!:H_e(0):\;\!\!\;\!\!:}
\def\noHm0{:\;\!\!\;\!\!:H_\mu(0):\;\!\!\;\!\!:}
\def\nof{:\;\!\!\;\!\!:}
\def\vect#1{{\bm #1}}
\def \r{\mathrm}
\def\Dia{\Diamond}\def\lab{\label}
\def\lan{\langle}\def\lar{\leftarrow}
\def\lf{\left}\def\lrar{\leftrightarrow}
\def\Lrar{\Leftrightarrow}\def\noi{\noindent}
\def\non{\nonumber}\def\ot{\times}
\def\pa{\partial}\def\ran{\rangle}
\def\rar{\rightarrow}\def\Rar{\Rightarrow}
\def\ri{\right}\def\ti{\tilde}
\def\we{\wedge}\def\wti{\widetilde}
\def\al{\alpha}\def\bt{\beta}\def\ga{\gamma}
\def\Ga{\Gamma}\def\de{\delta}\def\De{\Delta}
\def\ep{\epsilon}\def\ze{\zeta}\def\te{\theta}
\def\ka{\kappa}\def\la{\lambda}
\def\La{\Lambda}\def\si{\sigma}\def\Si{\Sigma}
\def\om{\omega}\def\Om{\Omega}
\def\AB{{_{A,B}}}
\def\CP{{_{C\!P}}}\def\T{{_{T}}}
\def\AB{{_{A,B}}}\def\mass{{_{1,2}}}
\def\flav{{e,\mu}}\def\1{{_{1}}}\def\2{{_{2}}}
\def\nof{:\;\!\!\;\!\!:}
\def\wQ{Q}
\def\wwQ{Q}

\def\I{{_{\rm{I}}}}\def\II{{_{\rm{II}}}}
\def\A{{_{A}}}\def\B{{_{B}}}

\def\rran{\ran\!\ran}
\def\llan{\lan\!\lan}

\title{A note on oscillating neutrino states in quantum field theory}
\author{Massimo Blasone\\ blasone@sa.infn.it}
%
\address{Dipartimento di Fisica, Universit\`a di Salerno, Via Giovanni Paolo II, 132 84084 Fisciano, Italy \& INFN Sezione di Napoli, Gruppo collegato di Salerno, Italy}

\author{Luca Smaldone\\ smaldone@ipnp.mff.cuni.cz}
%
\address{Faculty  of  Mathematics  and  Physics, Charles  University, V  Hole\v{s}ovi\v{c}k\'{a}ch  2, 18000  Praha  8,  Czech  Republic}

\maketitle
\begin{abstract}
In a recent paper (Eur.\ Phys.\ J.\ C {\bf 80}: 68 (2020).), a definition of oscillating neutrino states in quantum field theory was proposed. We show that such definition can be derived in a particular case of the Blasone--Vitiello approach, when mass vacuum is chosen as the physical vacuum. We discuss some problems of such an approach, which appears to be mathematically inconsistent and physically not acceptable. 
\end{abstract}

\section{Introduction}
Although many facts on neutrino oscillation physics are nowadays well established both theoretically \cite{giuntibook, NeutNotes,Ahluwalia:2004wf} and experimentally \cite{neutrinoexp} , the discussion on the correct treatment of flavor states in quantum field theory (QFT) permeated the last thirty years. The main approaches can be grouped as: \emph{external wave-packets} approach \cite{MixingReview} , the \emph{weak process states} approach \cite{GKL92} and \emph{Blasone-Vitiello (BV) approach} \cite{BlaVit95, qftmixing} . In the first one, flavor neutrino states are not explicitly constructed, because neutrinos are only regarded as internal lines of macroscopic Feynman diagrams. In the second one it is assumed that flavor neutrino states can be expanded as linear combination of mass eigenstates, where the amplitudes of such expansion depends on the production/detection process under consideration. In the third one, an Hilbert space of flavor states is explicitly constructed by means of a Bogoliubov (canonical) transformation. This \emph{flavor Fock space} is not the same as the Fock space of mass eigenstates, and flavor and mass representations turns out to be unitarily inequivalent representations of canonical anticommutation relations\cite{UMZ2} . 

In Ref.\cite{Tureanu:2019pui} it was proposed a universal, i.e. process independent, definition of oscillating neutrino states. This construction was based on the self-consistent method of Umezawa, Takahashi and Kamefuchi \cite{UTK} , originally based on the work of Nambu and Jona--Lasinio \cite{NJL} (see also Ref. \cite{DynMix}), where an appropriate boundary condition on the Hamiltonian fixes parameters of a Bogoliubov transformation, which maps all inequivalent choices for the physical Fock space. The author identifies the \emph{mass-Fock space} (see Ref.\cite{BlaVit95}) , as the physical one and consequently defines oscillating flavor states associated to flavor fields, by means of a Bogoliubov transformation which implements a mass shift. Clearly, within this method, the choice of the boundary conditions plays a crucial role \cite{DynMix} .

In this paper we show that the definition of oscillating neutrino states of Ref.\cite{Tureanu:2019pui} can be obtained as a particular case of the BV approach \cite{BlaVit95,qftmixing}, when the mass-Fock space is assumed to be the physical one and a particular choice of some parameters of the model is performed. In such a context, the construction of a flavor Fock space does not play any role. We also propose some arguments which show that these choices are physically inconsistent.

\section{Oscillating neutrino states form general Bogoliubov transformation}
Let us consider the Lagrangian density 
\be \label{mixlag}
\mathcal{L}(x)\ =\ \overline{\nu}(x) \, \lf(i \ga^\mu \pa_\mu \ - \ M_\nu \ri) \, \nu(x) \, ,
\ee
with
\be
\nu(x) \ = \
\bbm
\nu_e (x)\\ \nu_\mu (x)
\ebm \, , \qquad
M_\nu \ = \
\bbm
m_e & m_{e \mu} \\ m_{e \mu} & m_\mu
\ebm \, .
\ee
This can be diagonalized thanks to the transformation:
\bea
\label{PontecorvoMix}\nu_e (x) &=& \nu_1 (x)\cos\theta + \nu_2(x)\sin\theta \, , \\[2mm]
\nu_{\mu}(x) &=& -\nu_1(x)\sin\theta + \nu_2 (x)\cos\theta \, , \label{PontecorvoMix1}
\eea
with $\tan 2 \theta = 2 m_{e\mu}/(m_\mu-m_e)$. $\nu_e$ and $\nu_{\mu}$ are called flavor fields while $\nu_1$ and $\nu_2$ are the mass fields. The masses $m_1$ and $m_2$ are related to the original parameters as $m_e = m_1 \, c^2_\theta +  m_2 \, s_\theta^2$ and $m_\mu = m_1  s^2_\theta  +  m_2 c_\theta^2$, with $s_\theta \equiv \sin \theta$ and $c_\theta \equiv \cos \theta $.

Now $\nu_1$ and $\nu_2$ can be expanded as ($j=1,2$)
\bea 
\nu_j(x)  \ = \  \frac{1}{\sqrt{V}}\sum_{\mathbf{k},r} \,  e^{i{\bf k}\cdot {\bf x}} \, \lf[u_{\mathbf{k},j}^r \, \alpha_{\mathbf{k},j} \, e^{-i \, \om_{\G k,j} \, t}  + v_{-\mathbf{k},j}^r \, \beta_{-\mathbf{k},j}^{r\dagger} \, e^{i \om_{\G k,j} \, t}\ri] \, . \label{Fourierfield}
\eea
We can expand flavor fields in a similar way \cite{BlaVit95,qftmixing}:
\bea
\nu_\si(x)  =  \frac{1}{\sqrt{V}}\sum_{\mathbf{k},r} \,  e^{i{\bf k}\cdot {\bf x}} \, \lf[u_{\mathbf{k},\si}^r \, \alpha_{\mathbf{k},\si}(t) \, e^{-i \, \om_{\G k,\si} \, t}  + v_{-\mathbf{k},\si}^r \, \beta_{-\mathbf{k},\si}^{r\dagger}(t) \, e^{i \om_{\G k,\si} \, t}\ri], \, 
\si=e,\mu  , \label{Fourierfieldf}
\eea
where $\om_{\G k,\si} = \sqrt{|\G k|^2+\mu_\si^2}$ and $\mu_\si$  are mass parameters which have to be related with $m_1$ and $m_2$ by physical considerations \cite{Blasone:2011zz, Blasone:2018obn}. 
 Flavor creation and creation and annihilation operators have the general form\cite{FHY99} :
\bea \label{4x4Bog}  
  &&\hspace{-4mm} \left[\begin{tabular}{c} $\alpha^r_{\G k,e}$ \\ $\beta_{-\G k,e}^{r \dagger}$
\\$\alpha^r_{\G k,\mu}$ \\ $\beta_{-\G k,\mu}^{r \dagger}$ \end{tabular}\right]  
 = \left[\begin{array}{cccc}
c_\theta\, \rho^{\G k}_{e 1}& i \, c_\theta \,\lambda^{\G k}_{e 1} &
s_\theta \,\rho^{\G k}_{e 2}  &
i \, s_\theta \,\lambda^{\G k}_{e 2} \\
  i \, c_\theta \,\lambda^{\G k*}_{e 1} & c_\theta\, \rho^{\G k*}_{e 1} & i \, s_\theta
\,\lambda^{\G k*}_{e 2} & s_\theta \,\rho^{\G k*}_{e 2}
\\ 
 - s_\theta \,\rho^{\G k}_{\mu 1} & -i \, s_\theta \,\lambda^{\G k}_{\mu 1}& c_\theta
\,\rho^{\G k}_{\mu 2}
& i \, c_\theta \,\lambda^{\G k}_{\mu 2} \\ - i \, s_\theta \,\lambda^{\G k*}_{\mu 1} & -
s_\theta\,
\rho^{\G k*}_{\mu 1} &  i \, c_\theta\, \lambda^{\G k*}_{\mu 2}& c_\theta\, \rho^{\G k*}_{\mu 2}
\end{array}\right]\!
  \left[\begin{tabular}{c} $\al^r_{\G k,1}$ \\ $\bt_{-\G k,1}^{r\dagger}$ \\  $\al^r_{\G k,2}$ \\ $\bt_{-\G k,2}^{r\dagger} $\end{tabular}\right] \, . 
\eea
Here $\rho^\G k_{ab}=|\rho^\G k_{ab}|e^{i(\om_{\G k,a}-\om_{\G k,b})t}$, $\la^\G k_{ab}=|\la^\G k_{ab}|e^{i(\om_{\G k,a}+\om_{\G k,b})t}$, and
\begin{eqnarray}
 |\rho^\G k_{a b}| & \equiv & \cos\frac{\chi_a - \chi_b}{2}, \quad
|\lambda^\G k_{a b}| \ \equiv \  \sin\frac{\chi_a -
\chi_b}{2} \, , \non \\[2mm] \label{rholambda}
\chi_a & \equiv &  \cot^{-1}\lf[\frac{k}{m_a}\ri] \, ,  \qquad m_a, \, m_b \ = \ m_1, \, m_2, \, \mu_e, \, \mu_\mu\, .
\end{eqnarray}
The transformation \eqref{4x4Bog} is a canonical (Bogoliubov) transformation, which connects unitarily (and then physically) inequivalent representations of canonical anticommutation relations\cite{FHY99} : the \emph{mass} and the \emph{flavor} representation. 

We now show that definition of oscillating neutrino states of Ref.\cite{Tureanu:2019pui} can be incorporated in this scheme. In fact, setting $\mu_e=\mu_\mu=0$, one can define
\be \label{turneu}
|\nu^r_{\G k,\si}\ran_m \ \equiv \ \al^\dag_{\G k,\si}(0)|0\ran_{1,2} \, ,
\ee
where $|0\ran_{1,2}$ is the \emph{mass vacuum}, which is annihilated by $\al^r_{\G k,j}$, $\bt^r_{\G k,j}$. Explicitly, defining $\rho^\G k_{j} \ \equiv \ \cos\frac{\chi_j}{2} e^{-i \, \om_{\G k,j} \,t}$ and $\lambda^\G k_{j} \ \equiv \  \sin\frac{\chi_j}{2} e^{i \, \om_{\G k,j} \,t}$, we get:
\be \label{oscstate}
\bbm 
|\nu^r_{\G k,e}\ran_m \\[1mm] |\nu^r_{\G k,\mu}\ran_m 
\ebm 
\ = \ 
\bbm
c_\theta \, |\rho^{\G k}_1|  & s_\theta \, |\rho^{\G k}_2| \\[1mm]
-s_\theta \, |\rho^{\G k}_1|  & c_\theta \, |\rho^{\G k}_2|
\ebm
\,
\bbm
|\nu^r_{\G k,1}\ran \\[1mm] |\nu^r_{\G k,2}\ran 
\ebm \, .
\ee
$|\nu^r_{\G k,j}\ran \ = \ \al^{r,\dag}_{\G k,j}|0\ran_{1,2}$ are the mass eigenstates. This definition is actually identical to Eqs. (40),(41) of Ref.\cite{Tureanu:2019pui} . In the following we discuss such definition.
\section{Problematic aspects}
Adopting the choice of Ref.\cite{Tureanu:2019pui} , we consider the limit case $m_{e\mu}=0$, i.e. $\theta=0$ and then $m_e=m_1$ and $m_\mu=m_2$. In that case, flavor fields are expanded as free fields (cf. Eq.\eqref{Fourierfield}). If we set $\theta=0$ in Eq.\eqref{4x4Bog}, with $\mu_e=\mu_\mu=0$, we get
\bea
  \left[ \begin{tabular}{c} $\alpha^r_{\G k,e}$ \\ $\beta_{-\G k,e}^{r \dagger}$
\\$\alpha^r_{\G k,\mu}$ \\ $\beta_{-\G k,\mu}^{r \dagger}$ \end{tabular}\right]
= \left[\begin{array}{cccc}
\rho^{\G k}_1 & - i \, \lambda^{\G k}_{1} &
0  & 0
\\  -i \, \lambda^{\G k*}_{1} & \rho^{\G k*}_{1} & 0 & 0
\\ 0 & 0 & \rho^{\G k}_{2}
& - i \, \lambda^{\G k}_{2} \\  0 &
0 & - i \, \lambda^{\G k*}_{2}& \rho^{\G k*}_{2}
\end{array}\right]
\left[\begin{tabular}{c} $\al^r_{\G k,1}$ \\ $\bt_{-\G k,1}^{r\dagger}$ \\  $\al^r_{\G k,2}$ \\ $\bt_{-\G k,2}^{r\dagger} $\end{tabular}\right] \, .
 \label{4x4Tur1}
\eea
This is not compatible with the expansion \eqref{Fourierfield}. The only physical admissible representations are evidently those for which $\lim_{\theta \rightarrow 0}\mu_e=m_1$, $\lim_{\theta \rightarrow 0}\mu_\mu=m_2$, i.e. those which are connected with the identity \cite{Blasone:2011zz} . This can be also inferred looking at the limit of neutrino states \eqref{turneu}:
\be
\lim_{\theta \rightarrow 0} \, |\nu^r_{\G k,\si}\ran_m \ = \ |\rho^\G k_j| \, |\nu^r_{\G k,j}\ran \, , 
\ee
with $(\si,j)=(e,1),(\mu,2)$. This gives the expected result (i.e. the normalized mass state) only if $|\rho^\G k_j|=1$, i.e. in the trivial case $m_j \rightarrow 0$ considered in Ref.\cite{Tureanu:2019pui} . 

The choice of mass vacuum as physical vacuum could seem to be the most natural one, because it diagonalizes the Hamiltonian associated to the Lagrangian \eqref{mixlag}, which is the usual boundary condition to fix physical Fock space \cite{UTK} . However, in the case of mixing the situation is much more delicate \cite{DynMix} . 
To understand this point, consider a weak decay  $W^+\rightarrow e^+ + \nu_e$, which is associated with production or detection process. The relevant part of the effective Standard Model Lagrangian (after spontaneous symmetry breaking) is
$ \mathcal{L}_W=\mathcal{L}_0+\mathcal{L}_{int}$ with
\bea
&&\mbox{\hspace{-2mm}}{\cal L}_{0}  =  \overline{\nu}  \lf( i \ga_\mu \pa^\mu - M_{\nu} \ri)\nu \, + \, \overline{l} \lf( i \ga_\mu \pa^\mu - M_{l} \ri) l  \nonumber \, , \\[2mm]
 &&\mbox{\hspace{-2mm}}{\cal L}_{int}  =  \frac{g}{2\sqrt{2}}
\lf [ W_{\mu}^{+}\,
\overline{\nu}\,\gamma^{\mu}\,(1-\gamma^{5})\,l +
h.c. \ri] \, ,
\label{L-interact}
\eea
where $l = \lf[e 	\  \mu \ri]^T$, and
\bea
M_l\,=\,  \lf[\ba{cc}\tilde{m}_e &0 \\
0 & \tilde{m}_\mu\ea\ri] 	\,  .
\eea
The Lagrangian \eqref{mixlag} is just the neutrino part of $\mathcal{L}_0$, describing neutrino propagation.
The entire Lagrangian $\mathcal{L}_W$ is invariant under the global $U(1)$ transformations
$\nu \rightarrow e^{i \alpha} \nu$ and $l \rightarrow e^{i \alpha} l$
leading to the conservation of the total  flavor charge $Q_{l}^{tot}$ corresponding to the lepton-number conservation~\cite{BilPet} . This can be written in terms of the flavor charges for neutrinos and charged leptons
\be
Q_{l}^{tot} =  \sum_{\si=e,\mu} Q_\si^{tot}(t) \,,\quad   Q_{\si}^{tot} (t) = Q_{\nu_{\si}}(t) + Q_{\si}\,,
\ee
with
\bea
Q_{\si} & = &  \intx \,
:l_\si^{\dag}(x)l_\si(x): \,,\\[2mm] 
Q_{\nu_{\si}} (t) & = &  \intx \,
:\nu_{\si}^{\dag}(x)\nu_{\si}(x):\, , \quad \si=e,\mu \, .
 \label{QflavLept}
\eea
Note that $[\mathcal{L}_{0}({\bf x},t),Q_\si^{tot}(t)] \neq 0$. However, by observing that  $[\mathcal{L}_{int}({\bf x},t),Q_\si^{tot}(t)]=0$, we see that a neutrino flavor state is well defined in the production vertex as an eigenstate of the corresponding  flavor charge~\cite{Chargeei}. 

This corresponds to the fact that flavor of a neutrino is \emph{defined} by the flavor of the associated charged lepton \cite{BilGiu}: according to Ref. 20, when neutrinos are produced and detected, they ``\emph{carry identity cards}'', i.e. a definite flavor. Moreover, they ``\emph{can surreptitiously change them if given the right opportunity}'', i.e. during their propagation.

The oscillating neutrino states of Ref.\cite{Tureanu:2019pui} , i.e. the states \eqref{turneu}, spoil the defining property of flavor states\footnote{A similar problem was encountered also in Ref.\cite{BlaVit95} , where neutrino states were defined as in Eq.\eqref{turneu}, but choosing a different mass parametrization ($\mu_e=m_1$, $\mu_\mu=m_2$).}. Moreover, the oscillation probability, as defined in Ref.\cite{Tureanu:2019pui} 
\be
P_{\nu_\si \rightarrow \nu_\rho} (t)\ = \ |{}_m\lan \nu^r_{\G k,\si}(t)|\nu^r_{\G k,\rho}\ran_m|^2 \, , \quad \rho \neq \si \, .
\ee
with the time evolved (Schr\"odinger representation) neutrino state given by $|\nu^r_{\G k,\si}(t)\ran_m$ $\equiv$ $\al^\dag_{\G k,\si}(t)|0\ran_{1,2}$, leads to absurd consequences as:
\be
P_{\nu_\si \rightarrow \nu_\rho} (0) \ = \ \lf(c^2_\theta \, |\rho^{\G k}_1|^2+s^2_\theta \, |\rho^{\G k}_2|^2 \ri)^2\ \neq \ 1 \, .
\ee
In other words, there is a non-zero probability that an electronic neutrino at $t=0$ were not an electronic neutrino at the same time. The problem arises because $|\nu^r_{\G k,e}\ran_m$ and $|\nu^r_{\G k,\mu}\ran_m$ are not orthogonal.
\section{Conclusion}
In this paper we have shown that the definition of oscillating neutrino states introduced in Ref.\cite{Tureanu:2019pui} , can be seen as a particular case of BV formalism, when  $\mu_e=\mu_\mu=0$ and the mass vacuum is taken as physical vacuum. Both these choices are actually discussed: the first one brings to a discontinuity with the case $\theta=0$ (no-mixing), while the second one leads to a violation of lepton number conservation in the production/detection vertex, at tree level and to an oscillation probability formula with a bad-behaved initial condition. We thus conclude that the formalism of Ref. \cite{Tureanu:2019pui} leads to unphysical results.

\smallskip

Finally, we briefly reply to some criticisms presented in Ref.\cite{Tureanu:2020odo} :
\vspace{-2mm}
\begin{itemize}
\item
Flavor ladder operators defined in Eq.\eqref{4x4Bog} are not associated to positive and negative frequency modes. Therefore, in Ref. \cite{Tureanu:2020odo} it was claimed that the construction based on such operators is unphysical. Such flavor creation/annihilation operators are defined so that they create/annihilate modes with a definite \emph{flavor}, on the zero-flavor state, i.e. flavor vacuum. These have a completely different physical meaning with respect to usual (mass) ladder operators, which create/annihilate modes with definite \emph{energy} on the mass vacuum, which is an energy eigenstate. This difference is well known and does not imply any unphysical consequences \cite{Poi, Blasone:2018ktu}.
\item
In Ref. \cite{Tureanu:2020odo} it was stated that energy-conservation law would be violated once flavor vacuum were assumed as physical vacuum. This is not true: energy and flavor are incompatible variables, as explicitly shown in Ref. \cite{Blasone:2018ktu} , where a \emph{flavor--energy uncertainty relation} was derived, which establishes a bound on energy precision measurements. An explicit analysis of energy conservation, with Blasone--Vitiello approach, was performed in Ref. \cite{Lee} in the case of beta decay.
\item
In Ref. \cite{Tureanu:2020odo} it was stated that because of Coleman's theorem \cite{Miransky}, a theory based on flavor vacuum should prevent mixing, because lepton number for each family should be conserved. This is obviously false. The reason is that the proof of the Coleman's theorem requires Poincar\'e invariance of vacuum. However, flavor vacuum is \emph{not} Poincar\'e invariant. This point was discussed in detail in Ref.\cite{Poi} , in the bosonic case, where it was also shown that, despite this fact, flavor oscillation formula has the same form in all Lorentz frames \cite{Poi}. 
\end{itemize}
\section*{Acknowledgments}

The work of L.S. was supported by Charles University Research Center (UNCE/SCI/013).



\end{document}